\documentclass[twocolumn,nofootinbib,floatfix,10pt,aps]{revtex4-2}
\usepackage{amsmath}
\usepackage{amssymb}
\usepackage{wasysym}
\usepackage{graphicx}
\usepackage{color,soul}
\usepackage{physics}
\usepackage{siunitx}
\usepackage{dsfont}
\usepackage{multirow}
\usepackage{mathdots} 
\usepackage[english]{babel}
\usepackage{blindtext}
\usepackage[english,nomargin,inline,marginclue,draft]{fixme}
\pdfpageheight\paperheight
\pdfpagewidth\paperwidth
\usepackage[colorlinks,linkcolor=blue,anchorcolor=blue,citecolor=blue,urlcolor=blue]{hyperref}
            
\fxusetheme{colorsig}
\FXRegisterAuthor{cg}{acg}{CG}  
\FXRegisterAuthor{th}{ath}{\color{blue}TH}  
\FXRegisterAuthor{ib}{aib}{\color{red}IB} 
\FXRegisterAuthor{sh}{ash}{\color{cyan}SH} 
\FXRegisterAuthor{db}{adb}{\color{green}DB} 
\FXRegisterAuthor{ps}{aps}{PS}
\makeatletter
\renewcommand*\FXLayoutInline[3]{%
  {\@fxuseface{inline}\ignorespaces{\color{fx#1}[#3: #2]}}}
\makeatother

\long\def\symbolfootnote[#1]#2{\begingroup%
\def\thefootnote{\fnsymbol{footnote}}\footnotetext[#1]{#2}\endgroup}

\def\nobreakbefore{%
  \relax\ifvmode\else
    \ifhmode
      \ifdim\lastskip > 0pt\relax
        \unskip\nobreakspace
      \else 
        \nobreakspace
      \fi
    \fi
  \fi
}
\let\oldcite\cite
\renewcommand\cite{\nobreakbefore\oldcite}

\begin{document}

\title{Spectral fingerprint recognition of mobile signals via Rydberg atomic sensors}

\author{Yi-Ming Yin$^{1,2,3}$}
\thanks{These authors contributed equally to this work.}
\author{Ming-Yao Cui$^{4}$}
\thanks{These authors contributed equally to this work.}
\author{Ling-Yu Du$^{5}$}
\thanks{These authors contributed equally to this work.}
\author{Jin-Hang Fang$^{6}$}
\author{Luan-Yi Wei$^{6}$}
\author{Bang Liu$^{1,2,6}$}
\author{Li-Hua Zhang$^{1,2,6}$}
\author{Kai Wang$^{5}$}
\author{Kai-Bin Huang$^{4}$}
\email{huangkb@hku.hk}
\author{Dong-Sheng Ding$^{1,2,3,6}$}
\email{dds@ustc.edu.cn}

\affiliation{$^1$Key Laboratory of Quantum Information, University of Science and Technology of China, Hefei, Anhui 230026, China.}
\affiliation{$^2$Synergetic Innovation Center of Quantum Information and Quantum Physics, University of Science and Technology of China, Hefei, Anhui 230026, China.}
\affiliation{$^3$Hefei National Laboratory, University of Science and Technology of China, Hefei, 230088, China.}
\affiliation{$^4$Department of Electrical and Computer Engineering, The University of Hong Kong, Hong Kong SAR, China.}
\affiliation{$^5$Electronic Engineering Institute, National University of Defense Technology, Hefei, Anhui 230037, China}
\affiliation{$^6$Davinci Quantum Technology Co., Ltd, Hefei, Anhui 230026, China}
\date{\today}

\begin{abstract}
Rydberg-atom sensors have emerged as a promising quantum technology for broadband radio-frequency detection owing to their intrinsic calibration, wide frequency coverage, and high sensitivity. However, their potential for intelligent identification of real-world wireless signals remains largely unexplored. Here, we demonstrate a quantum-enabled framework for mobile signal classification based on Rydberg electromagnetically induced transparency spectroscopy and deep learning. Three types of wireless emissions, including cellular voice call, Wi-Fi, and 5G signals from three commercial smartphone brands, are measured to construct a spectral-fingerprint dataset. A convolutional neural network trained on this dataset achieves an initial baseline classification accuracy of 93.44\%. To compensate for spectral drifts induced by temporal and environmental variations, a progressive fine-tuning strategy is employed to continuously adapt the pretrained model, enabling robust classification across diverse deployment scenarios. Furthermore, continuous real-time monitoring is demonstrated under dynamically varying wireless conditions, including cellular voice call, Wi-Fi, and their coexistence. The proposed system achieves an overall frame-level recognition accuracy of 87.67\% over a 60-s measurement. These results establish a pathway toward intelligent quantum radio-frequency sensing and highlight the potential of Rydberg-atom platforms for real-time wireless signal awareness and spectrum monitoring.
\end{abstract}

\maketitle

\section{Introduction}
The rapid expansion of wireless communication networks has led to increasingly complex radio-frequency (RF) environments, posing significant challenges for signal identification and spectrum management\cite{1391031, 4796930}. In particular, distinguishing between smartphone emissions such as cellular voice calls, Wi-Fi, and 5G signals has become an important requirement for modern wireless monitoring systems\cite{8054694, 9171707}. Conventional approaches typically rely on electronic receivers combined with digital signal processing and machine learning algorithms to extract discriminative features from measured waveforms\cite{Like2009, 5399351}. In parallel, radio-frequency fingerprinting methods have been explored for leveraging subtle physical-layer characteristics of wireless devices for identification tasks\cite{7902125, 10436939, 10012861, 10415087, 10538532, Bitar2017wireless, Kulin2018spectrum, OShea2018overAir, Merchant2018RF, Shi2019dynamic}. Although effective in controlled settings, these methods often face challenges in maintaining robust performance under varying environmental conditions owing to receiver calibration errors, hardware complexity, and susceptibility to electromagnetic interference.

In recent years, Rydberg-atom-based quantum sensors have emerged as a promising platform for RF field detection. Owing to their exceptionally large electric dipole moments and polarizabilities, Rydberg atoms exhibit strong interactions with external electromagnetic fields over a broad frequency range. Through electromagnetically induced transparency (EIT)\cite{Foot1998, Fleischhauer2005, AbiSalloum2010, Liao2020}, ac Stark shifts\cite{AutlerTownes1955, Foot1998, Boh2007}, and Autler–Townes splitting, incident RF signals can be directly mapped onto optical transmission spectra, enabling all-optical readout of electromagnetic information. Compared with conventional metallic antennas\cite{Sedlacek2012}, this atom-based transduction provides broadband and frequency-selective RF sensing, with the operating frequency range accessible across different RF bands through appropriate atomic transitions and sensing configurations\cite{Holloway2014,
Holloway2017, Meyer2020, Meyer2021, ZHANG2024, PhysRevA.90.043419, Cui2025RARE}. This cross-band sensing capability\cite{10.1063/1.4890094, Cui2025RARE}, together with intrinsic self-calibration\cite{Holloway2014, Holloway2017}, high sensitivity\cite{Bason2010, LiuSensitive, Meyer2020, Meyer2021, Ding2022, ZHANG2024, ZHANG20241515, Wang2025ArcRF, Wang2026, PhysRevA.90.043419, Jing2020superheterodyne, Cui2025RARE}, and broadband operation, makes Rydberg receivers attractive for sensing complex and spectrally diverse RF environments. These advantages have stimulated extensive research on quantum RF sensing, communication reception, field metrology, and signal demodulation\cite{PhysRevApplied.18.014033, cpl_42_5_053201, 10238372, zhang2024floquet, Zhang:24, Simons2021rydberg, Robinson2021angle, Cai2023phase, Cui2025RARE, Nan2026ultrabroadband}.

\begin{figure*}[htb]
    \centering\includegraphics[width=2\columnwidth]{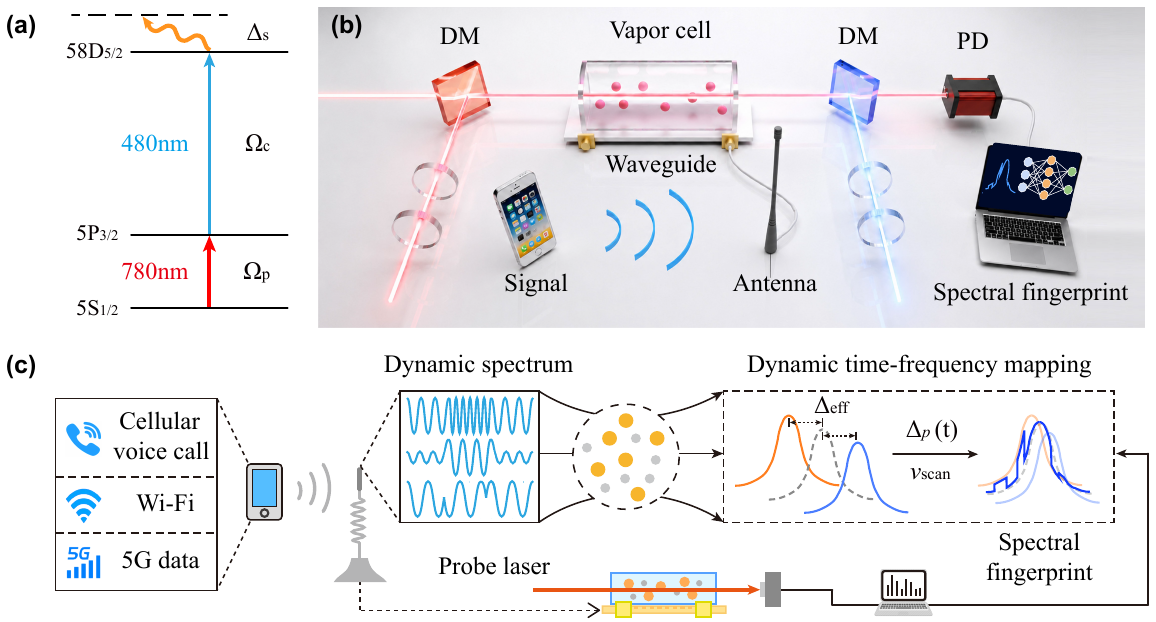}
    \caption{\textbf{Illustration of the setup and physical diagram.} 
    (a) Energy-level diagram. The probe (780nm) and coupling (480nm) laser drive the transitions among the ground state $5S_{1/2}$, the intermediate state $5P_{3/2}$, and the Rydberg state $58D_{5/2}$. In the presence of phone signals, the Rydberg level $58D_{5/2}$ exhibits an energy shift $\Delta_{s}$ governed by the ac Stark effect. 
    (b) Experimental setup. Wireless signals are captured by an antenna, guided through a waveguide, and radiated onto the Rb vapor cell. The EIT transmission of the probe beam is detected by a photodetector (PD) to obtain spectral fingerprints. 
    (c) Schematic workflow for signal recognition. Different transmission modalities (cellular voice calls, Wi-Fi, and 5G data) generate distinct time-varying RF waveforms, which induce different perturbations of the Rydberg level and imprint characteristic spectral fingerprints on the EIT spectra for identification.}
    \label{fig1}
\end{figure*}

Despite this remarkable progress, most Rydberg-atom-based studies have primarily focused on RF field measurements, frequency estimation, communication reception, and signal demodulation\cite{PhysRevApplied.18.014033, cpl_42_5_053201, 10238372, zhang2024floquet, Zhang:24}. Whether the rich spectral responses of Rydberg atoms can be exploited as robust physical fingerprints for intelligent classification of real-world wireless signals remains largely unexplored \cite{Liu2022, Yin:26, WangPD}. Moreover, practical deployment of quantum sensing systems requires reliable operation under temporal drifts, environmental perturbations, and hardware-induced spectral variations, yet the impact of such distribution shifts on intelligent signal recognition has received little attention. Addressing these challenges is essential for transforming Rydberg-atom sensors from precision measurement devices into intelligent platforms for autonomous electromagnetic perception.

In this work, we demonstrate an adaptive, quantum-enabled framework for wireless signal classification based on Rydberg-atom EIT spectroscopy and deep learning. A comprehensive dataset comprising more than 10,000 spectral fingerprints generated by voice call, Wi-Fi, and 5G emissions from three smartphone brands is constructed for model training and evaluation. A convolutional neural network is employed to learn discriminative representations directly from the measured EIT spectra, achieving an initial baseline classification accuracy of 93.44\%. To address performance degradation caused by environmental and temporal variations, we introduce a progressive fine-tuning strategy that incrementally adapts the pretrained model, effectively recovering the classification accuracy to above 92\% in moderate regimes and to 89.44\% even in the most severely distorted scenario. Finally, real-time experiments are conducted to validate the capability of the proposed approach for dynamic wireless signal identification, including cellular voice call, Wi-Fi, and mixed signals, achieving an overall frame-level recognition accuracy of 87.67\% over a 60-s measurement. These results establish a robust, adaptive paradigm for intelligent quantum radio-frequency sensing, bridging the gap between laboratory-grade quantum metrology and real-world electromagnetic intelligence.

\begin{figure*}[htb]
    \centering\includegraphics[width=1.6\columnwidth]{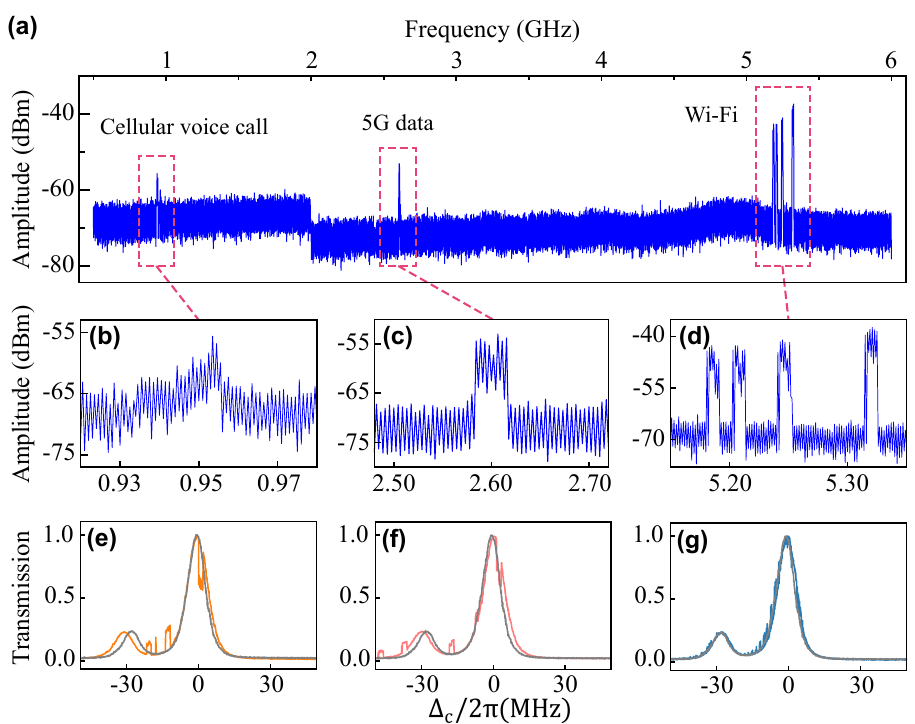}
    \caption{\textbf{Construction of spectral fingerprints.} 
    (a) Broadband frequency spectrum containing representative cellular voice call, 5G, and Wi-Fi signals. 
    (b–d) Zoomed-in views of the corresponding spectral regions, showing their distinct spectral structures and modulation characteristics. 
    (e–g) Rydberg-EIT spectra measured under exposure to the three signals. Relative to the reference spectrum (gray), each signal induces a characteristic modification of the EIT line shape due to the unique spectral content of each signal. The resulting atomically encoded responses exhibit signal-dependent features that form reproducible optical spectral fingerprints, providing the basis for subsequent signal identification and classification. }
    \label{fig2}
\end{figure*}

\section{Physical model}
The experimental configuration and the corresponding atomic energy-level scheme are illustrated in Figs.~\ref{fig1}(a,b). A ladder-type EIT configuration is implemented in a rubidium vapor cell, where a weak probe laser and a strong coupling laser coherently drive the transitions from the ground state $|g\rangle$ ($\rm5S_{1/2}$) to the intermediate state $|e\rangle$ ($\rm5P_{3/2}$), and subsequently to the Rydberg state $|r\rangle$ ($\rm58D_{5/2}$). Owing to their exceptionally large electric dipole moments and polarizabilities, Rydberg atoms provide an ultrasensitive optical interface capable of converting incident wireless electromagnetic fields into measurable changes of the EIT transmission spectrum.

Unlike monochromatic microwave radiation, practical smartphone emissions, including 4G LTE, 5G NR, and Wi-Fi signals, are broadband and nonstationary. The incident wireless field can be represented as a superposition of dynamically modulated spectral components,
\begin{equation}
    E_{\rm MW}(t)
    =
    \frac{1}{2}
    \sum_k
    A_k(t)
    e^{i[\omega_k t+\phi_k(t)]}
    +\mathrm{c.c.},
    \label{eq:wireless_field}
\end{equation}
where $A_k(t)$, $\omega_k$, and $\phi_k(t)$ denote the instantaneous amplitude, angular frequency, and phase of the $k$th component, respectively. In an operational wireless network, these quantities are governed not only by the encoded information but also by protocol-level processes, including power control, resource-block allocation, packet retransmission, and frequency switching. Consequently, even when the carrier band remains unchanged, the spectral content of a practical wireless signal evolves continuously in time. To characterize this nonstationarity, we introduce the local time-dependent spectrum $S(\nu,t)$ through the short-time Fourier transform (STFT),
\begin{equation}
\begin{aligned}
    \widetilde{E}(\nu,t)
    &=
    \int_{-\infty}^{\infty}
    E_{\rm MW}(\tau)
    w(\tau-t)
    e^{-i2\pi\nu\tau}
    \,d\tau,\\
    S(\nu,t)
    &=\frac{c\epsilon_0}{2}
    \left|
    \widetilde{E}(\nu,t)
    \right|^2 ,
\end{aligned}
\label{eq:stft}
\end{equation}
where $w(\tau-t)$ is a finite temporal window of duration $T_w$ centered at $t$. The spectral density $S(\nu,t)$ is in units of $\mathrm{W\,m^{-2}\,Hz^{-1}}$ and describes the local distribution of wireless power over frequency within a short time interval $T_w$. 

\begin{figure*}[htb]
    \centering\includegraphics[width=2\columnwidth]{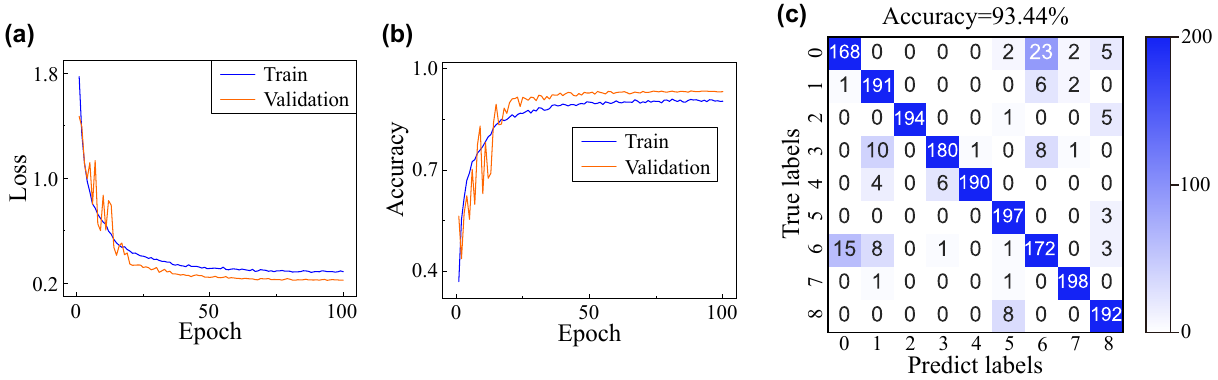}
    \caption{\textbf{Recognition performance of wireless communication signals.} 
    (a,b) Evolution of training and validation loss and accuracy over epochs for the CNN classifier, with training (blue) and validation (orange) datasets. 
    (c) Confusion matrix after 100 training epochs, showing an overall classification accuracy of 93.44\%. }
    \label{fig3}
\end{figure*}

In our experimental configuration, the target wireless signals occupy the 1--6~GHz frequency range and are far detuned from the sensitive dipole-allowed transitions involving $\ket{58D_{5/2}}$. The RF field therefore perturbs the Rydberg level through off-resonant ac Stark coupling to the neighboring Rydberg manifold. Under a locally stationary approximation, the instantaneous shift of the Rydberg state can be modeled as~\cite{simons2018electromagnetically}
\begin{equation}
\begin{aligned}
    \Delta_{\rm eff}(t)
    &=
    \sum_{f\neq r}
    \frac{
        e^2\nu_{fr}^{3}
        \left|
        \mathbf{n}\cdot
        \bra{f}\hat{\mathbf r}\ket{r}
        \right|^2
    }{
        \hbar h c\epsilon_0
    }
    \int_{\nu_{\min}}^{\nu_{\max}}
    \frac{
        S(\nu,t)
    }{
        \nu^2\left(\nu^2-\nu_{fr}^2\right)
    }
    \,d\nu ,
\end{aligned}
\label{eq:dynamic_stark}
\end{equation}
where $\ket{r}=\ket{58D_{5/2}}$ is the target Rydberg state and $\ket{f}$ denotes an electric-dipole-coupled state in the neighboring Rydberg manifold. Here, $\nu_{fr}=(E_f-E_r)/h$ is the transition frequency, $\mathbf n$ is the polarization unit vector of the RF field, $\hat{\mathbf r}$ is the position operator, and $e$, $h$, $\hbar$, $c$, and $\epsilon_0$ have their usual meanings. The integration range $[\nu_{\min},\nu_{\max}]$ covers the wireless-frequency band incident on the atoms.
Equation~(\ref{eq:dynamic_stark}) shows that the instantaneous Rydberg-state shift is determined by a frequency-dependent weighted integral of the local wireless spectrum $S(\nu,t)$. Thus, temporal variations in the spectral content of different communication signals are converted into corresponding time-dependent perturbations of the Rydberg level. The resulting shift causes the EIT resonance to move along the detuning axis, with its temporal evolution determined by the incident wireless signal. These signal-dependent atomic responses subsequently give rise to distinguishable spectral fingerprints for wireless-signal recognition.

The motion of the EIT resonance is recorded by continuously scanning the probe-laser detuning at an approximately constant rate,
\begin{equation}
    \Delta_p(t)
    =
    \Delta_{p,0} + 
    v_{\rm scan}t,
    \label{eq:probe_scan}
\end{equation}
where $\Delta_{p,0}$ is the initial probe detuning and $v_{\rm scan}$ is the scan velocity. The scan establishes a one-to-one mapping between the acquisition time and the horizontal coordinate of the recorded spectrum. Neglecting the finite response time of the EIT system, the optical transmission measured at time $t$ is modeled  as
\begin{equation}
    I_{\rm meas}(\Delta_p)
    =
    I_{\rm EIT}
    \left[
        \Delta_p(t)
        -
        \Delta_{\rm eff}
        \left(t
        \right)
    \right],
    \label{eq:mapping}
\end{equation}
where $I_{\rm EIT}(\Delta_p)$ denotes the unperturbed EIT transmission profile. Notably, the recorded spectrum is not a snapshot of a stationary resonance. Instead, each point on the spectral line is acquired at a different time while the center of the EIT resonance is itself moving.

For sufficiently small shifts, $|\Delta_{\rm eff}|\ll\Gamma_{\rm EIT}$, Eq.~(\ref{eq:mapping}) can be linearized as
\begin{equation}
\begin{aligned}
    I_{\rm meas}(\Delta_p)
    \simeq\,
    &I_{\rm EIT}(\Delta_p)-
    \Delta_{\rm eff}
    \left(t
    \right)
    \frac{dI_{\rm EIT}}{d\Delta_p}.
\end{aligned}
\label{eq:linearized_mapping}
\end{equation}
The resulting distortion is therefore determined by the product of the instantaneous shift and the local slope of the intrinsic EIT line. For example, a red shift increases the sampled transmission on the positive-slope wing of the EIT resonance, whereas the same shift produces the opposite response on the negative-slope wing. Rapid variations in $\Delta_{\rm eff}(t)$ are thus projected by the laser scanning into localized peaks, dips, and spike-like features in the recorded EIT spectrum, establishing a direct time-to-frequency mapping of the atomic response. Consequently, the measured EIT spectrum represents a dynamic optical projection of the incident wireless waveform rather than the response to a stationary monochromatic field. The resulting line-shape distortions constitute reproducible atomic spectral fingerprints that encode characteristic information about different wireless communication protocols.

\begin{figure*}[htb]
    \centering\includegraphics[width=1.8\columnwidth]{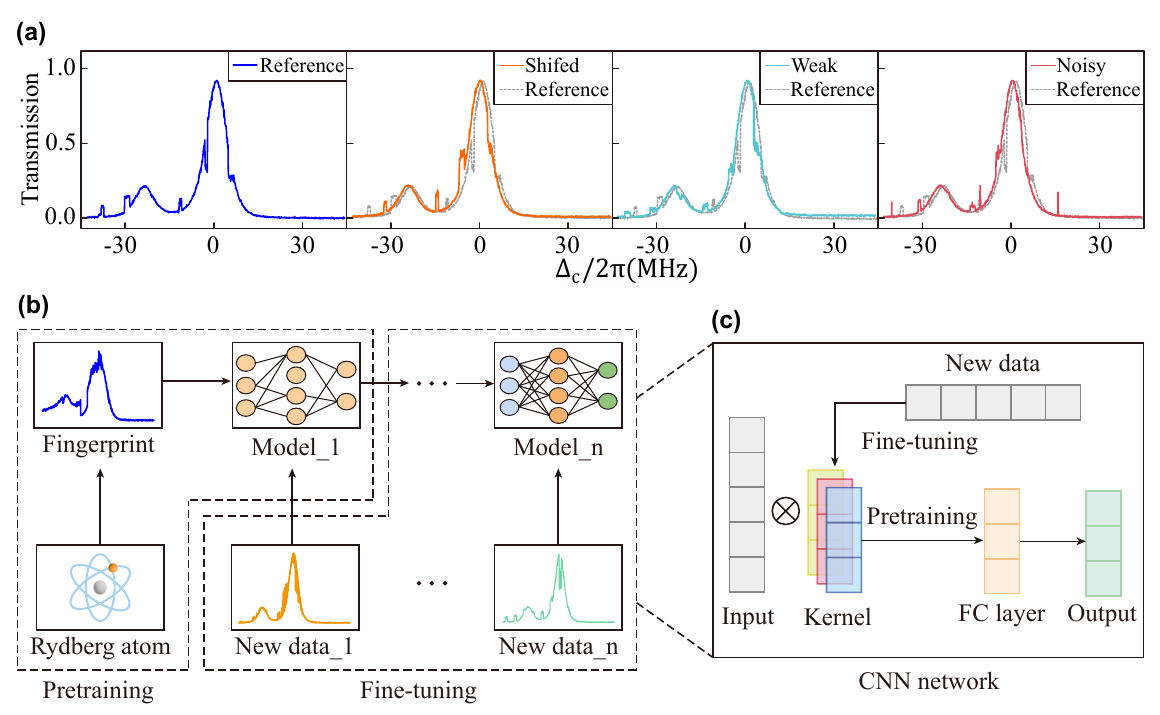}
    \caption{\textbf{Spectral fingerprint variations and the model adaptation framework.}
    (a) Representative atomic spectral fingerprints, including a reference fingerprint (blue) and three distorted variants exhibiting spectral shift (orange), weakened features (cyan), and added noise (red).
    (b) Schematic illustration of the model adaptation procedure, consisting of a pretraining stage based on reference spectral fingerprints and a subsequent fine-tuning stage using newly acquired datasets.
    (c) CNN architecture employed for fingerprint classification, where pretrained feature representations are further refined during fine-tuning to adapt to spectral variations in newly acquired data. }
    \label{fig4}
\end{figure*}

\section{Results}

\subsection{The construction and recognition of fingerprints}
To evaluate the capability of the Rydberg-atom sensor for wireless signal perception, we first investigate whether different communication signals can be encoded into distinguishable optical fingerprints. Three representative wireless communication modes, including a cellular voice call, a 5G data transmission, and a Wi-Fi signal, were collected and applied to the atomic sensing system. As shown in Fig.~\ref{fig2}(a), these signals occupy different regions within the 0–6 GHz frequency range and exhibit distinct spectral characteristics. The enlarged spectra in Figs.~\ref{fig2}(b–d) further reveal their different bandwidths, spectral structures, and modulation features. The cellular voice call signal features a narrow, localized peak centered near 0.95 GHz, indicative of traditional narrowband channel allocations. Conversely, the 5G data signal exhibits a well-defined, flat-topped allocation around 2.60 GHz, characteristic of modern orthogonal frequency-division multiplexing (OFDM) or wideband code-division multiple access architectures. The Wi-Fi transmission presents a highly dynamic, multi-channel burst profile operating above 5.20 GHz, which reflects its packet-based, high-throughput nature.

When the incident RF fields interact with the Rydberg atoms, their spectral information is transferred to the optical domain through perturbation of the EIT resonance. The resulting EIT spectra are shown in Figs.~\ref{fig2}(e–g). Compared with the reference spectrum recorded in the absence of external RF signals, each communication signal produces a unique modification of the EIT line shape. The cellular voice-call signal introduces asymmetric distortions and localized spectral features around the main resonance. The 5G signal produces a broader redistribution of the transmission profile accompanied by pronounced fine structures. In contrast, the Wi-Fi signal generates dense modulation features superimposed on the central EIT peak. These signal-dependent perturbations arise from the different spectral compositions and modulation characteristics of the incident wireless signals. Consequently, the atomic system converts RF information into reproducible optical responses, forming characteristic spectral fingerprints that preserve the intrinsic properties of the original signals.

\begin{figure*}[htb]
    \centering\includegraphics[width=2\columnwidth]{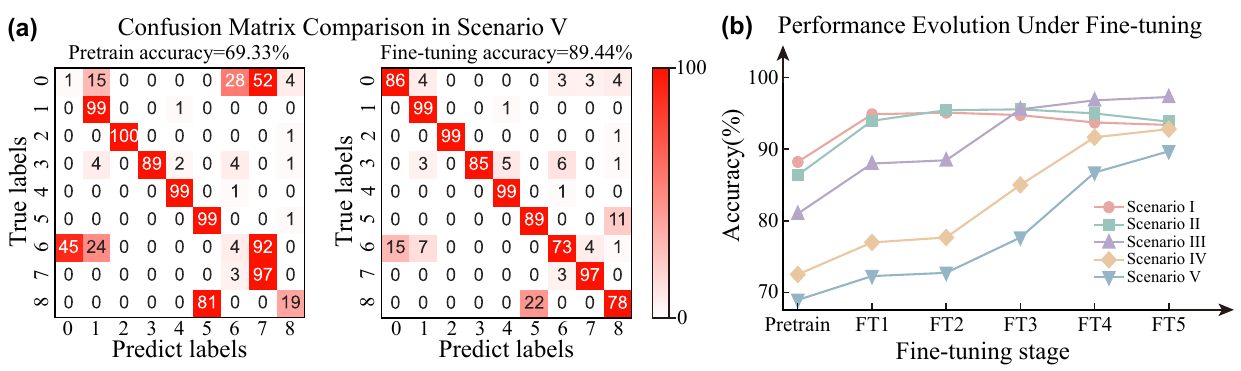}
    \caption{\textbf{Fine-tuning enhances classification performance under spectral variations.} 
    (a) Comparison of confusion matrices for Scenario V before (left) and after (right) fine-tuning, with the overall classification accuracy improving from 69.33\% to 89.44\%.
    (b) Classification accuracies across five representative scenarios during progressive model adaptation, from the pretrained model (Pretrain) through five successive fine-tuning stages (FT1–FT5).
}
    \label{fig5}
\end{figure*}

To assess the discriminability of these fingerprints, signals collected from three smartphone brands under the three communication modes were used to construct a nine-category dataset. The resulting spectral fingerprints were then classified using a CNN model. Figures.~\ref{fig3}(a-b) present the training and validation loss and accuracy during the optimization process. Both the training and validation curves exhibit stable convergence, indicating that the network successfully learns representative features from the atomic fingerprints. After 100 training epochs, the model achieves a classification accuracy of 93.44\%, as shown by the confusion matrix in Fig.~\ref{fig3}(c). The strong diagonal dominance demonstrates that most samples are correctly assigned to their corresponding categories. These results confirm that the proposed Rydberg-atom spectral fingerprints contain sufficient discriminative information for wireless signal recognition and establish a foundation for intelligent spectrum sensing based on quantum sensors.

\subsection{Adaptation of the recognition model to spectral variations}
In practical deployments, Rydberg-atom-based spectral fingerprints are inevitably affected by environmental fluctuations, hardware instabilities, and measurement noise, leading to distortions in both spectral position and line shape. These variations effectively induce a mismatch between training and testing distributions, causing a degradation in classification performance when models trained on pristine data are directly applied to unseen conditions. To address this issue, we investigate model robustness under progressively increasing spectral variations and develop a sequential fine-tuning strategy for continuous adaptation.

Figure.~\ref{fig4}(a) illustrates representative spectral distortions relative to the pristine Rydberg-EIT fingerprint, including frequency shift, feature attenuation, and noise corruption. These perturbations not only modify peak positions and amplitudes but also obscure fine spectral structures that encode discriminative information, thereby increasing the difficulty of feature extraction in the learned representation space. To mitigate this effect, a transfer-learning-based adaptation framework is adopted, as shown in Fig.~\ref{fig4}(b). Starting from a pretrained CNN model, successive fine-tuning is performed using datasets corresponding to increasing distortion levels (Scenario I → V), yielding a cascade of adapted models (FT1–FT5). The network architecture is illustrated in Fig.~\ref{fig4}(c), where convolutional layers extract hierarchical spectral features and fully connected layers perform classification.

The adaptation effect is first illustrated using the most challenging case (Scenario V) in Fig.~\ref{fig5}(a). Before fine-tuning, the confusion matrix shows severe off-diagonal misclassification, reflecting strong class overlap induced by spectral distortions, with an overall accuracy of 69.33\%. After sequential fine-tuning, misclassifications are substantially suppressed and the matrix exhibits substantially stronger diagonal dominance, indicating improved separability in the feature space. Correspondingly, the classification accuracy increases to 89.44\%, demonstrating effective realignment of decision boundaries under strongly shifted spectral distributions.

The evolution of classification performance under sequential adaptation is summarized in Fig.~\ref{fig5}(b). The pretrained model exhibits progressive degradation as spectral variations increase, with accuracy decreasing from 88.00\% (Scenario I) to 69.33\% (Scenario V), reflecting increasing distribution shift severity. After the first fine-tuning stage, performance is significantly recovered across all scenarios, indicating rapid adaptation to the initial domain shift. As sequential adaptation proceeds (FT2–FT5), further improvements are observed in moderately and severely distorted regimes, with final accuracies reaching 96.78\%, 92.44\%, and 89.44\% for Scenarios III–V, respectively. For mildly perturbed cases (Scenarios I and II), performance rapidly saturates after early adaptation and remains stable, with only minor fluctuations within a narrow range (~1–2\%), suggesting that the model preserves robust representations once the dominant distribution shift has been compensated.

\begin{figure*}[htb]
    \centering\includegraphics[width=2.0\columnwidth]{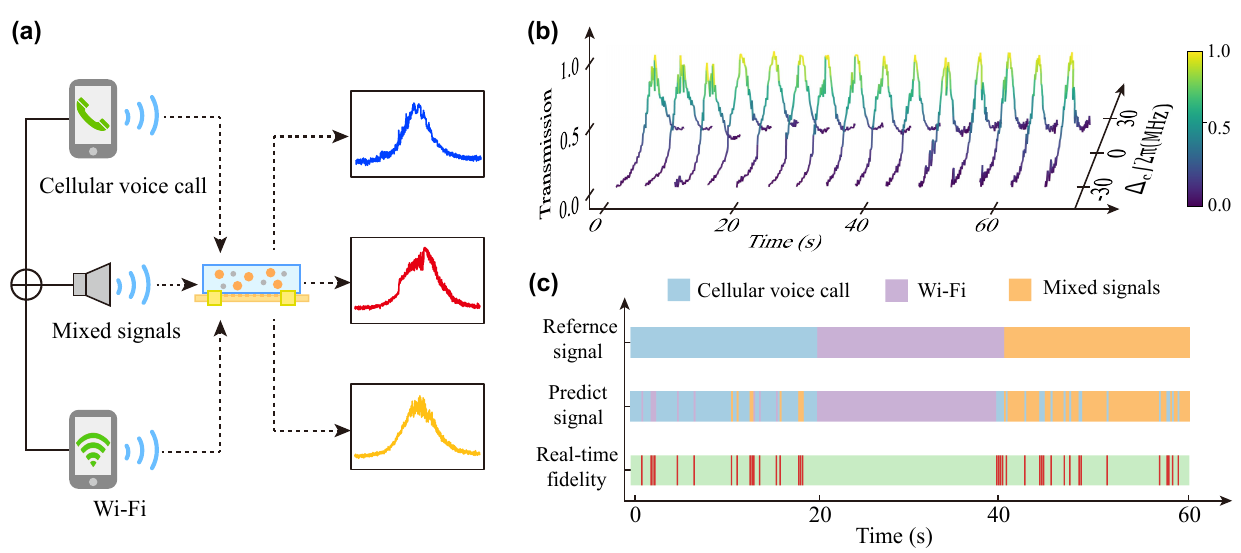}
    \caption{\textbf{Real-time recognition of dynamic wireless signals via continuous atomic spectral fingerprinting.}
    (a) Schematic of the real-time sensing process for cellular voice call, Wi-Fi, and mixed signals. The incident wireless signals are coupled to the Rydberg atomic sensor, where their characteristic electromagnetic features are converted into distinct optical spectral fingerprints.
    (b) Continuous stream of Rydberg-EIT spectra recorded over 60 s as the incident signal condition is sequentially switched among cellular voice call, Wi-Fi, and mixed signals.
    (c) Comparison between the reference signal sequence and the corresponding real-time predictions, together with the frame-by-frame recognition fidelity. The proposed system achieves an overall real-time classification accuracy of 87.67\%.
    \label{fig6}
}
\end{figure*}

\subsection{Real-time recognition of dynamic wireless signals}
Having demonstrated the capability of the Rydberg-atom sensor to generate discriminative spectral fingerprints, we further investigate its performance under continuously varying wireless conditions. In contrast to the static classification experiments, real-world wireless environments may involve signal appearance, disappearance, and coexistence, requiring the sensing system to continuously track changes in the incident electromagnetic environment.

As illustrated in Fig.~\ref{fig6}(a), three representative signal conditions were considered, including a cellular voice call, a Wi-Fi transmission, and a mixed signal in which the two signals coexist. Each signal condition produces a characteristic perturbation of the Rydberg-EIT spectrum, resulting in distinguishable atomic spectral fingerprints. The measured spectra were continuously acquired and processed frame by frame to evaluate the capability of the system for dynamic signal recognition.

Figure.~\ref{fig6}(b) shows the time-resolved stream of Rydberg-EIT spectra recorded over 60 s. The incident signal condition was sequentially switched among cellular voice call, Wi-Fi, and mixed signals, leading to corresponding temporal variations in the measured atomic spectra. The continuous evolution of the spectral stream demonstrates that the Rydberg sensor can track changes in the surrounding wireless environment while preserving signal-dependent spectral features.

The corresponding recognition results are presented in Fig.~\ref{fig6}(c), where the reference signal sequence is compared with the predicted sequence generated by the trained CNN. The prediction follows the major temporal transitions among the three signal conditions, including the coexistence of cellular and Wi-Fi signals. Recognition errors are more pronounced around signal-transition regions and during the mixed-signal interval, where the measured spectra exhibit more complex temporal variations. Despite these deviations, the predicted sequence maintains good temporal correspondence with the reference sequence throughout the 60-s measurement. The overall frame-level classification accuracy reaches 87.67\%, demonstrating the capability of the Rydberg-atom sensing system to perform continuous real-time recognition of dynamically varying and coexisting wireless signals.

\section{Discussion}
In this work, we demonstrate a quantum-enabled framework that synergizes Rydberg-atom spectroscopy with deep learning for intelligent wireless signal recognition. By converting cellular voice, Wi-Fi, and 5G wireless signals into structured optical spectral fingerprints, we establish a direct link between communication protocols and their corresponding atomic responses. The resulting fingerprints can be classified by a convolutional neural network with an accuracy of 93.44\%, while a progressive fine-tuning strategy effectively compensates for performance degradation induced by environmental fluctuations and system instabilities. Furthermore, real-time experiments demonstrate continuous recognition of cellular voice call, Wi-Fi, and mixed signals, achieving an overall accuracy of 87.67\% over a 60-s measurement.

This study introduces a paradigm shift in RF perception. Unlike conventional receiver architectures that typically require frequency-dependent RF front ends and electronic frequency-conversion chains, our approach utilizes the intrinsic nonlinear interaction between RF fields and Rydberg atoms to perform an all-optical, physics-based transformation of electromagnetic data. The resulting fingerprints arise from the complex interplay of dynamic Stark shifts, multi-frequency dipole couplings, and the temporal statistics of the incident signals. The success of our model adaptation strategy suggests that these atomically encoded features are highly discriminative and remain robust against distribution shifts, highlighting the potential of the atomic medium as a high-dimensional physical feature extractor.

While this work establishes the foundation for intelligent quantum RF sensing, several opportunities for future exploration emerge. Expanding the dataset to include a broader spectrum of communication standards and diverse transmitter configurations will be critical to assessing the system’s scalability. Moreover, beyond sequential fine-tuning, the integration of advanced continual-learning algorithms could further enhance long-term adaptation with minimal reliance on labeled calibration data. Finally, transitioning toward high-speed optical interrogation and integrated, low-latency inference hardware will be instrumental in enabling fully autonomous operation in increasingly dense electromagnetic environments.

In summary, we have shown that Rydberg-atom platforms transcend their traditional role as quantum RF detectors and can function as intelligent sensing nodes capable of autonomous signal recognition. By bridging quantum sensing and machine intelligence, the proposed framework establishes a pathway toward intelligent quantum RF perception. These results open new opportunities for spectrum monitoring, wireless signal intelligence, electromagnetic situational awareness, and future cognitive radio systems.

\section{Method}
\subsection{Dynamic response of Rydberg-EIT coherence to time-varying wireless waveforms}
To quantitatively describe the dynamic optical response of the Rydberg-EIT system to incident wireless waveforms, we employ the time-dependent density-matrix formalism. The evolution of the driven three-level atomic ensemble is governed by the Liouville-von Neumann master equation,
    \begin{equation}
        \dot{\rho}(t)=-i[H(t),\rho(t)]+\mathcal{L}(\rho),
    \end{equation}
where $\rho(t)$ is the atomic density matrix and $\mathcal{L}(\rho)$ denotes the Lindblad superoperator accounting for spontaneous decay and coherence dephasing.

The probe laser continuously scans across the EIT resonance with a constant scanning velocity,
    \begin{equation}
        \Delta_p(t) = \Delta_{p,0} + v_{\rm scan} t,
    \end{equation}
where $\Delta_{p,0}$ is the initial probe detuning and $v_{\rm scan}$ is the scan velocity. During the scan, the incident wireless waveform introduces a time-dependent perturbation to the Rydberg resonance through the effective shift $\Delta_{\rm eff}(t)$ described by Eq.~(\ref{eq:dynamic_stark}). Under the off-resonant and quasi-static approximation adopted in the present experiment, the RF-induced interaction with neighboring Rydberg states is incorporated into this effective Stark shift. Therefore, no additional microwave-interaction term is introduced in the Hamiltonian. The Hamiltonian of the driven three-level system in the rotating frame can thus be written as
    \begin{equation}
    \begin{aligned}
        H(t)
        ={}&
        -\Delta_p(t)|e\rangle\langle e| \\
        &-
        [\Delta_p(t)+\Delta_c-\Delta_{\rm eff}(t)]
        |r\rangle\langle r| \\
        &+
        \frac{1}{2}
        \left(
        \Omega_p|g\rangle\langle e|
        +
        \Omega_c|e\rangle\langle r|
        +\mathrm{H.c.}
        \right),
    \end{aligned}
    \label{eq:hamiltonian}
    \end{equation}
where $\Omega_p$ and $\Omega_c$ denote the probe- and coupling-field Rabi frequencies, respectively, and $\Delta_c$ is the coupling-laser detuning.

Under the weak-probe approximation ($\Omega_p\ll\Omega_c$), the atomic population remains predominantly in the ground state ($\rho_{gg}\approx1$). The optical response is therefore determined by the probe coherence,
    \begin{equation}
        \dot{\rho}_{eg}(t)=-\left[\gamma_{eg}-i\Delta_p(t)\right]
        \rho_{eg}(t)+\frac{i}{2}\Omega_p+\frac{i}{2}\Omega_c\rho_{rg}(t),
    \end{equation}
where $\rho_{rg}(t)$ denotes the Rydberg coherence continuously perturbed by the incident wireless waveform through Eq.~(\ref{eq:hamiltonian}).

The instantaneous probe transmission through the vapor cell follows the Beer-Lambert law,
    \begin{equation}
        I_{\rm snap}(t)=I_0\exp\left[-kL\,{\rm Im}
        \left(\rho_{eg}(t)\right)\right],
    \end{equation}
where $k$ is the probe wavenumber, $L$ is the interaction length of the vapor cell, and $I_0$ is the incident probe intensity.

Because the probe frequency is continuously scanned, each probe detuning corresponds uniquely to one acquisition time,
    \begin{equation}
        t = \frac{\Delta_p-\Delta_{p,0}}{v_{\rm scan}}.
    \end{equation}

Accordingly, the experimentally measured EIT spectrum is obtained by directly mapping the instantaneous optical transmission onto the probe-frequency axis,
    \begin{equation}
        I_{\rm meas}(\Delta_p)=I_{\rm snap}\left(
        \frac{\Delta_p}{v_{\rm scan}}\right),
    \end{equation}
which is equivalent to the time-to-frequency mapping introduced in the Physical Model.

Substituting the instantaneous transmission into the above relation yields
    \begin{equation}
        I_{\rm meas}(\Delta_p)=I_0\exp\left[-kL\,{\rm Im}\left(\rho_{eg}\left(\frac{\Delta_p}{v_{\rm scan}}\right)\right)\right].
    \end{equation}

Therefore, the measured EIT spectrum preserves both the intrinsic quantum response of the Rydberg-EIT system and the temporal evolution of the incident wireless waveform. The slowly varying envelope of the wireless signal is continuously encoded onto the probe-frequency axis during laser scanning, resulting in reproducible spectral fingerprints that can be directly utilized for wireless signal identification.

\subsection{Deep learning model and continual adaptation strategy}
To classify atomic spectral fingerprints corresponding to different wireless communication modalities, we employed a lightweight one-dimensional convolutional neural network (1D-CNN). The network was designed to provide high classification accuracy while maintaining low computational complexity, thereby facilitating rapid adaptation to evolving spectral conditions.

Prior to training, each spectral fingerprint was represented as a one-dimensional amplitude sequence. To ensure consistent input dimensions, all samples were either truncated or padded with trailing zeros to a fixed target length of 1018 points. Each waveform was subsequently normalized using sample-wise $Z$-score standardization:
    \begin{equation}
        x_i' = \frac{x_i - \mu}{\sigma + \varepsilon},
    \end{equation}
where $x_i$ denotes the original amplitude, $\mu$ and $\sigma$ represent the mean and standard deviation of the individual waveform, respectively, and $\varepsilon = 10^{-8}$ is a small regularization constant to prevent numerical instability.

\begin{figure}[htb]
    \centering\includegraphics[width=\columnwidth]{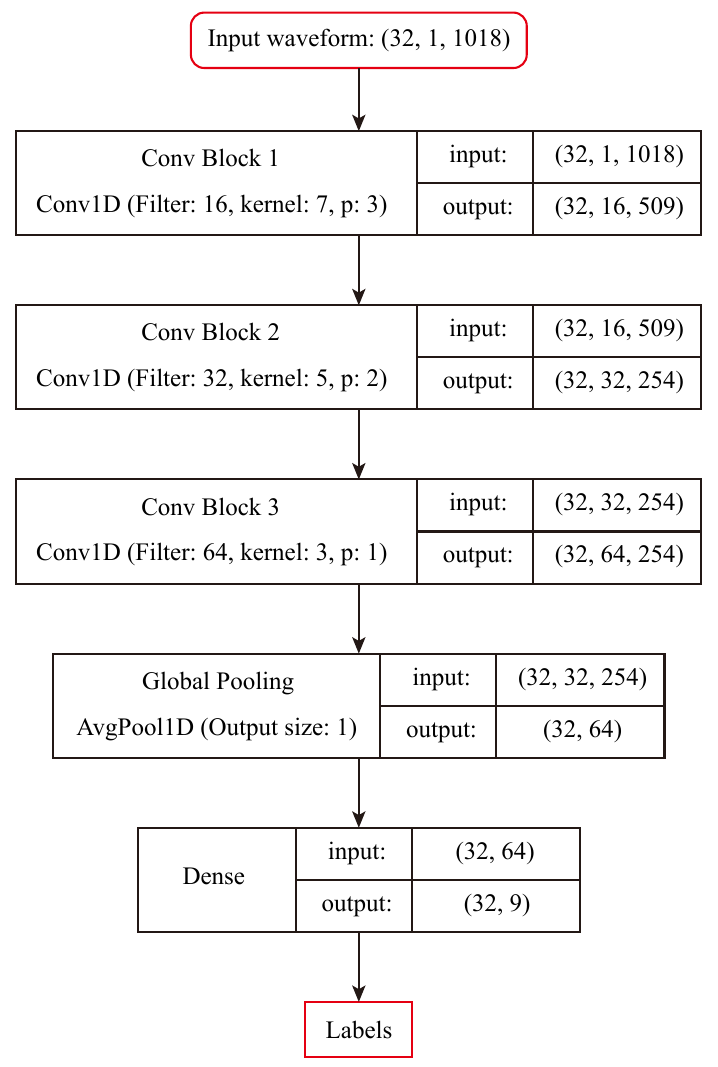}
    \caption{\textbf{Architecture of our 1D convolutional neural network.} 
    The network consists of three convolutional blocks with progressively increasing channel numbers, followed by global average pooling and a fully connected classification layer. The input spectral waveform is transformed into a compact feature representation and subsequently mapped to the corresponding communication-signal category.
}
    \label{fig7}
\end{figure}

The neural network architecture is illustrated in Fig.~\ref{fig7}. The model consists of three sequential convolutional stages. The first stage applies a 1D convolution with 16 output channels and a kernel size of 7, followed by batch normalization, rectified linear unit (ReLU) activation, and max pooling. The second stage employs a 32-channel convolution with a kernel size of 5 and an identical normalization–activation–pooling structure. The third stage uses 64 convolutional filters with a kernel size of 3, followed by batch normalization and ReLU activation. The forward convolution operation is expressed as:
    \begin{equation}
        z^{(c)}(n) = \sum_{m=0}^{k-1} x(n+m) \cdot W_m^{(c)},
    \end{equation}
where $W^{(c)}$ denotes the convolution kernel associated with the $c$-th output channel, and $k$ is the kernel length.

To generate a fixed-dimensional feature representation independent of the input signal length, the final convolutional feature maps were processed using adaptive global average pooling:
    \begin{equation}
        v_c = \frac{1}{T}\sum_{t=1}^{T}Z_{c,t},
    \end{equation}
where $Z_{c,t}$ denotes the activation of channel $c$ at temporal position $t$, and $T$ is the temporal dimension of the feature map. The resulting latent feature vector $\mathbf{v} \in \mathbb{R}^{64}$ was passed through a dropout layer with a penalty probability of 0.3 to mitigate overfitting, followed by a fully connected layer to produce the raw class logits:
    \begin{equation}
        \mathbf{y} = \mathbf{W}\mathbf{v} + \mathbf{b}.
    \end{equation}

Posterior class probabilities were subsequently obtained via the standard softmax function:
    \begin{equation}
        \hat{y}_i = \frac{\exp(y_i)}{\sum_{j=1}^{K}\exp(y_j)},
    \end{equation}
where $K$ denotes the total number of distinct communication categories encoded within the dataset.

The network parameters were optimized by minimizing the categorical cross-entropy loss objective function:
    \begin{equation}
        \mathcal{L}_{\rm CE} = -\sum_{i=1}^{K} y_i \log(\hat{y}_i),
    \end{equation}
where $y_i \in \{0,1\}$ is the one-hot encoded ground truth index. 

During the initial foundational pre-training stage, the network parameters were initialized with PyTorch defaults and optimized using the Adam optimizer for 100 epochs with a batch size of 32 and an initial learning rate of $\eta_0 = 10^{-3}$. The comprehensive dataset was randomly partitioned into independent training and validation subsets using stratified sampling with an 80:20 ratio to rigorously preserve pristine class distributions. To enhance generalization and prevent redundant optimization post-convergence, a \emph{ReduceLROnPlateau} learning rate scheduler monitored the validation loss trajectory. The learning rate was dynamically halved whenever the validation loss failed to decrease for three consecutive epochs:
    \begin{equation}
        \eta \leftarrow 0.5 \cdot \eta.
    \end{equation}
Following the pre-training horizon, the optimized backbone parameters yielding the peak validation performance were locked and serialized as the initialization state for subsequent tasks.

To accommodate gradual spectral variations induced by shifting experimental environments or measurement conditions, we adopted an incremental, sequential fine-tuning strategy. Specifically, the model configurations optimized on the preceding dataset served as the direct warm-start state for the consecutive adaptation deployment. During this fine-tuning phase, all network weights remained fully trainable to allow holistic adaptation, while the learning rate was scaled down by an order of magnitude to a conservative value of $\eta_{\rm FT} = 10^{-4}$ for a brief regime of 10 epochs. This paradigm ensures the efficient transfer and preservation of historically learned robust spectral topologies while preventing catastrophic forgetting under progressively evolving target data distributions.

The entire deep learning framework was implemented using the PyTorch library and executed on a high-performance GPU-accelerated computing platform. Model selection was governed by the validation cross-entropy minimum, and the corresponding optimal weights were saved for downstream multi-modal evaluation and field deployment.

\bibliographystyle{apsrev4-2}
\bibliography{ref}  

\end{document}